\documentclass[aps, prl, reprint, superscriptaddress, floatfix]{revtex4-2}

\usepackage{graphicx} 
\usepackage{braket}
\usepackage{physics}
\usepackage{amsmath}
\usepackage{amssymb}
\usepackage{mathtools}
\usepackage{hyperref}
\usepackage[capitalize]{cleveref}
\usepackage{xcolor}
\usepackage{bm}

\begin{document}

\title{Universality of Stochastic Control of Quantum Chaos with Measurement and Feedback}

\author{Andrew A. Allocca}
\affiliation{Department of Physics and Astronomy, Louisiana State University, Baton Rouge, Louisiana 70803, USA}
\affiliation{Center for Computation and Technology, Louisiana State University, Baton Rouge, Louisiana 70803, USA}

\author{Devesh K. Verma}
\affiliation{Department of Physics and Astronomy, Louisiana State University, Baton Rouge, Louisiana 70803, USA}

\author{Sriram Ganeshan}
\affiliation{Department of Physics, City College, City University of New York, New York, New York 10031, USA}
\affiliation{CUNY Graduate Center, New York, New York 10031, USA}

\author{Justin H. Wilson}
\affiliation{Department of Physics and Astronomy, Louisiana State University, Baton Rouge, Louisiana 70803, USA}
\affiliation{Center for Computation and Technology, Louisiana State University, Baton Rouge, Louisiana 70803, USA}

\begin{abstract}
We investigate universal features of measurement-and-feedback control of quantum chaotic dynamics by examining the quantum Arnold cat map, a paradigmatic model of quantum chaos. 
Inspired by probabilistic control of classical chaos, our protocol stochastically alternates between intrinsic instability and engineered control operations that steer trajectories toward a target point. 
Simulation of exact quantum dynamics and a semiclassical truncated Wigner approximation reveal universal properties of the cat map's control transition.
To further characterize this universality, we introduce the inverted harmonic oscillator as an analytically tractable effective model of instability. 
By integrating numerical simulations, a semiclassical Fokker-Planck description, and a direct spectral analysis of the stochastic quantum channel, we identify quantum signatures absent in classical limits.
The close agreement between quantum simulation, truncated Wigner approximation, and inverted oscillator analysis shows that universal features of the transition are set by uncertainty-limited quantum fluctuations and are insensitive to genuine quantum interference.
\end{abstract}

\maketitle

\textit{Introduction}---Order emerges from chaotic dynamics across a range of scales, from microscopic quantum systems to macroscopic classical, and even socioeconomic, processes.
In many-body systems, local interactions are sufficient to induce ordering.
Bird flocks and fish schools provide familiar classical examples~\cite{vicsek1995novel, toner1995long}, superconductivity and superfluidity represent analogous quantum phenomena, and within statistical mechanics, ordered states emerge through phase transitions.

Beyond passive emergence, order can be engineered from chaotic dynamics using both deterministic procedures, e.g. the Ott-Grebogi-Yorke method~\cite{Ott1990}, and probabilistic approaches, involving the stochastic intervention of a control map with probability $p$ into chaotic dynamics~\cite{Antoniou1996,Antoniou1997,Antoniou1998}.
In this scheme the control map and chaotic dynamics share a periodic orbit, unstable for the latter and stable for the former, which becomes a global attractor above a critical control rate $p_c$.
This is a random multiplicative process where distance from the control point is either amplified by a factor $e^{\kappa}$ by chaotic dynamics or contracted by a factor $e^{-\gamma}$ by control; this has been studied in turbulent systems \cite{Friedrich1997,Naert1997,Benzi2003,Durrive2020} and market dynamics \cite{Levy1994,Levy1995,Levy1996,Frisch1997,Laherrere1998,Takayasu2000,Friedrich2000,Sornette2001}.
These ideas have also recently gained resonance in quantum systems, where measurement-conditioned feedback operations can stabilize states and steer dynamics~\cite{Iadecola2023,Herasymenko2023,Buchhold2022,Ravindranath2022,ODea2022,Friedman2022,Piroli2023,Sierant2022,Sierant2023,LeMaire2024,Milekhin2024,Allocca2024,Ravindranath2025,Gerbino2025}. 

In this Letter, we study stochastic measurement-and-feedback control of quantized chaotic maps, focusing on the quantum Arnold cat map~\cite{Arnold1968}. 
Using exact quantum dynamics and the truncated-Wigner-approximation (TWA), we extract universal properties of the quantum control transition. 
Since essential features of the transition are determined by the local dynamics near the control point, it can be effectively described by an inverted harmonic oscillator (IHO)\footnote{The IHO arises broadly across physics \cite{Barton1986,Gentilini2015,Betzios2016,Dalui2019,Hegde2019,Subramanyan2021}.}. \nocite{Barton1986,Gentilini2015,Betzios2016,Dalui2019,Hegde2019,Subramanyan2021} 
Focusing on this model, we use numerical simulations of Gaussian-state evolution, a semiclassical Fokker-Planck (FP) analysis augmented by quantum uncertainty, and exact eigenoperators of the stochastic quantum channel, to characterize the transition and the controlled phase. 

Despite lacking a compact phase space, genuine quantum chaos, or quantum interference, the IHO shows excellent agreement with controlled cat-map dynamics.
Using the cat map as a paradigm, this indicates that for 2D chaotic dynamics the essential physics of \emph{quantum} stochastic control is governed locally near the unstable fixed point.
The measurement-and-feedback protocol suppresses wave-packet spreading and self-interference, effectively pushing the Ehrenfest time to infinity and rendering interference largely irrelevant for controllability, consistent with the broader role of continuous monitoring in the emergence of classical chaotic dynamics~\cite{Bhattacharya2000}.

\textit{Classical control}---We begin with a brief illustration of classical control of chaos.
Take $\mathbf{r}(t)$ to be the small displacement from an unstable fixed point or periodic orbit in a 2D phase space, where $t$ is a discrete time index. 
There is always a choice of coordinates where chaotic and control maps act on $\mathbf{r}(t)$ as $\mathcal{S}=\mathrm{diag}(e^\kappa,e^{-\kappa})$ and $\mathcal{C}=\mathrm{diag}(e^{-\gamma},e^{-\gamma})$ respectively, with $\kappa,\gamma>0$ parametrizing their respective strengths ($\det \mathcal{S} = 1$ due to being area preserving).
The behavior of the first component $r_1$ determines controllability of the system: chaos is controlled when the Lyapunov exponent $\overline{\ln\abs{r_1(t)}}<0$ as $t\to\infty$, where $\overline{(\cdots)}$ represents averaging over stochastic trajectories.
The average of this quantity can be performed analytically, resulting in a critical control rate \cite{Antoniou1998,Prasad2026},
\begin{equation} \label{eq:pc}
    p_c = \frac{\kappa}{\kappa+\gamma},
\end{equation}
above which the system is controlled, and below which chaos prevails.

\textit{Quantized chaotic map}---We use the Arnold cat map~\cite{Arnold1968} as a paradigmatic example of 2D chaos. 
Classically, this map transforms the unit square as
\begin{equation}
    \mathbf{r}(t+1) = \begin{pmatrix}
        2 & 1 \\
        1 & 1 \\
    \end{pmatrix}\mathbf{r}(t) \mod 1,
\end{equation}
and has fixed point $\mathbf{r}_0 = 0$ and strength $\kappa = 2\ln(\frac{1+\sqrt{5}}{2})$.
To quantize this map we promote the two coordinates to noncommuting quantum operators: $\mathbf{r}=(r_1\,\,\, r_2)\to\hat{\mathbf{r}}=(\hat{x}\,\,\,\hat{p})$, with $[\hat{x},\hat{p}] = i\hbar$.
The cat map can be implemented as unitary evolution on these operators with $\hat{U}_\mathrm{cat} = e^{-i\hat{p}^2/(2\hbar)}e^{i\hat{x}^2/(2\hbar)}$.
We take discrete position eigenstates $\ket{x_n=n/N}$ and momentum eigenstates $\ket{p_m=m/N}$, $n,m=0,1,\dots,N-1$.
With $\ket{x_n} = \frac{1}{N}\sum_m e^{2\pi i nm/N}\ket{p_m}$ and unit-square phase space, $\hbar=\frac{1}{2\pi N}$~\cite{Supplement,Hannay1980}.
The continuum limit $N\to\infty$ is therefore the classical limit. 

Because unitary quantizations arise from area-preserving maps~\cite{ArnoldWeinstein1989,DeBievreDegliEsposti1998}, implementing non-area-preserving control requires measurement.
We do this with a positive operator-valued measure (POVM) defined by Kraus operators.
To construct them, we introduce a basis via $[2 - \cos(2\pi \hat x) - \cos(2\pi \hat p)]\ket{n_c} = \epsilon_{n_c}\ket{n_c}$ such that $\ket{0}$ is the controlled state (fixing the phases of the basis by the harmonic oscillator convention, see \cite{Supplement}).
This allows us to construct Kraus operators
\begin{equation} \label{eq:ctrlKraus}
    \hat K_m(\theta) = \frac{(-i)^m}{\sqrt{m!}}\sin^m(\theta) (\cos(\theta))^{\hat{a}^\dagger \hat{a}} \hat{a}^m,
\end{equation}
where $\theta\in[0,\pi/2]$ parametrizes the strength of control, $\hat a = \sum_{n_c=0}^{N_\mathrm{ctrl}-1} \sqrt{n_c} \ket{n_c-1}\bra{n_c}$ where we take $N_\mathrm{ctrl} = N/2$, and complete the POVM with a projector on the remaining basis states. 
This control channel acts on $\hat{x}$ and $\hat{p}$ as a rescaling by $\cos\theta$, so classical and quantum control strengths $\gamma$ and $\theta$ are related as $\cos\theta=e^{-\gamma}$.

Both quantum channels (cat map and control) can be simulated with the TWA \cite{Polkovnikov2010}.
In the TWA, dynamics are classical with quantum fluctuations sampled from the initial Wigner function $W_0(x,p) = \frac1{\pi \hbar} e^{-(x^2+p^2)/\hbar}$, while control induces noise $(x,p)\mapsto \cos\theta(x,p)+\bm{\eta}$ with $\bm{\eta}$ Gaussian of variance $\frac{\hbar}{2}(1-e^{-2\gamma})$ (see \cite{Supplement}).
We use the late-time, trajectory-averaged squared overlap with the control state, as an order parameter for control, $\bar\rho_{00} = \overline{\ev{\hat \rho(t\to\infty)}{0}}$ where $\hat\rho(t)$ is the density matrix, which we can compute via the TWA under the full channel. 
This formulation neglects interference and follows from the linear approximation in \cref{eq:control_channel}, so control-POVM nonlinearities are neglected as well.

\begin{figure}
    \centering
    \includegraphics[width=0.87\columnwidth]{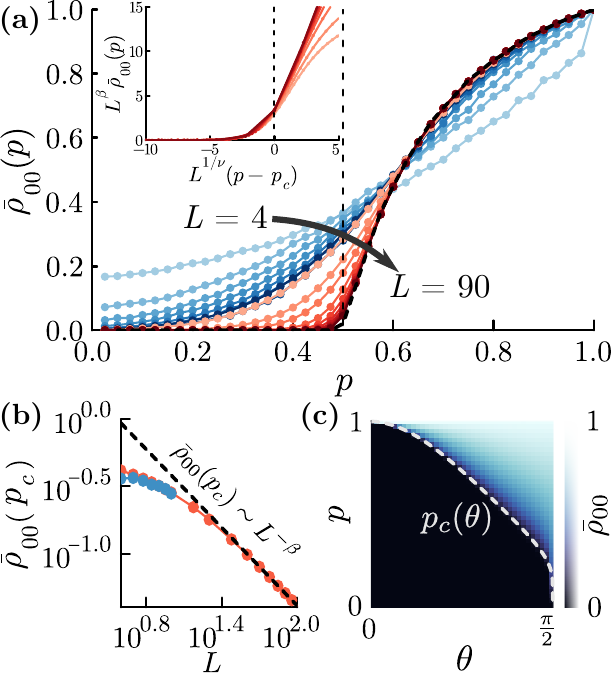}
    \caption{(a) Order parameter $\bar{\rho}_{00}(p)$ from exact cat-map simulations ($L=\log_2N=4$--$10$, blue) and TWA ($L=10$--$90$, red); inset: TWA collapse near $p=p_c$ with $\nu=1$ and $\beta=0.97$. (b) $\bar{\rho}_{00}(L)$ at $p_c$ for exact (blue) and TWA (red); dashed line: $\bar{\rho}_{00}\propto L^{-\beta}$ with $\beta=0.97$. (c) TWA $\bar{\rho}_{00}$ at $L=90$ vs $p,\theta$, consistent with the phase boundary $p_c(\theta)$ (white dash); $\hbar=\frac{1}{2\pi N}=\frac{1}{2^{L+1}\pi}$.}
    \label{fig:catmapfig}
\end{figure}

In \cref{fig:catmapfig}(a) we show $\bar{\rho}_{00}(p)$ for finite $\hbar$, where the transition is \emph{rounded} into a crossover but sharpens as $\hbar\rightarrow 0$ (i.e., $N\rightarrow\infty$).
Furthermore, we see this crossover smoothly connect the quantum data for $L\equiv\log_2N = 4$--$10$ to the TWA for $L=10$--$90$; the two agree quantitatively for $L=10$ ($N=1024$)~\cite{Supplement}.
The control strength $\theta$ is chosen so that $p_c=1/2$ as calculated by \cref{eq:pc}. 
Scaling collapse near $p=p_c$ gives correlation length exponent $\nu\approx 1$, and examining $\bar{\rho}_{00}$ as a function of $L$ at $p_c$, shown in \cref{fig:catmapfig}(b), gives the order parameter scaling exponent $\beta\approx 1$.
Figure \ref{fig:catmapfig}(c) shows $\bar\rho_{00}$ obtained with the TWA for $L=90$ for all $p$ and $\theta$, and clearly identifies the boundary between controlled and uncontrolled phases, agreeing with the expected classical result for $p_c$.

The success of the TWA in capturing the behavior of the quantum cat map's control transition indicates that quantum interference effects are largely irrelevant, and suggests that the transition admits a universal description. 
Motivated by this, we now isolate the dynamics near the unstable fixed point.

\textit{Inverted harmonic oscillator}---The saddle-point structure near an unstable fixed point is modeled by the IHO, 
\begin{equation} \label{eq:IHOHam}
    \hat{H} = \frac{\hat{p}^2}{2} - \frac{\Omega^2\hat{x}^2}{2},
\end{equation}
where $\Omega$ controls the strength of the potential. 
Time evolution under this Hamiltonian $\hat{U}(t) = \mathrm{exp}(-i\hat{H}t/\hbar)$ generates a single-mode squeezing operation and a Heisenberg-picture quantum channel
\begin{equation} \label{eq:UChannel}
    \mathbb{S}[\hat{\mathcal{O}}] = \hat{U}^\dagger(t)\,\hat{\mathcal{O}}\,\hat{U}(t).
\end{equation} 
The action of this channel for time $t_S$ models the effect of a quantum-chaotic map near the fixed point with strength $\kappa = \Omega\, t_S$. 
The operators $\hat{v}_\pm \equiv \sqrt{\Omega/2}\left(\hat{x}\pm\hat{p}/\Omega\right)$ grow or decay as $\mathbb{S}[\hat{v}_\pm] = e^{\pm \kappa} \hat{v}_\pm$, and all operators $(\hat{v}_+)^m(\hat{v}_-)^n$ are eigenvectors of the channel; $\expval{\hat{v}_\pm}$ are analogs of the components of the classical $\mathbf{r}$ above. 

As for the quantum cat map we can define ladder operators $\hat{a}$, $\hat{a}^\dagger$ and number states $\ket{n}$.
The vacuum $\ket{0}$ is the unstable fixed point of IHO evolution. 
Control of the IHO can be implemented by introducing an ancilla mode with its own ladder operators $\hat{b},\hat{b}^\dagger$ and number states $\ket{n}_b$.
Initializing the ancilla in its vacuum $\ket{0}_b$, coupling it to the IHO, then measuring and resetting it back into the vacuum state yields Kraus operators $K_m(\theta) = \bra{m}_b e^{-i \theta(\hat{a}^\dagger \hat{b} + \hat{a}\hat{b}^\dagger)} \ket{0}_b$ which have exactly the same form as \cref{eq:ctrlKraus}.
The Heisenberg-picture quantum channel for control is
\begin{equation} \label{eq:KChannel}
    \mathbb{C}[\hat{\mathcal{O}}] = \sum_{m=0}^\infty \hat{K}_m^\dagger(\theta)\,\hat{\mathcal{O}}\,\hat{K}_m(\theta).
\end{equation}
All normally ordered products of $\hat{a}$ and $\hat{a}^\dagger$ are eigenvectors of this channel, $\mathbb{C}[(\hat{a}^\dagger)^k \hat{a}^l] = \cos^{k+l}(\theta)\, (\hat{a}^\dagger)^k \hat{a}^l$.
As for \cref{eq:ctrlKraus}, $\mathbb{C}$ scales linear operators by $\cos\theta=e^{-\gamma}$. 

The full stochastic evolution of a generic operator acting on the IHO is the weighted sum of these two channels
\begin{equation} \label{eq:fullChannel}
    \hat{\mathcal{O}}(t+1) = \mathbb{T}[\hat{\mathcal{O}}(t)] = (1-p)\mathbb{S}_\kappa[\hat{\mathcal{O}}(t)] + p\,\mathbb{C}_\gamma[\hat{\mathcal{O}}(t)],
\end{equation}
where channels are labeled by their respective exponential parameters.

\textit{Quantum dynamics and Gaussian state evolution}---The dynamics of $\expval{\hat{v}_\pm}$ under $\mathbb{S}_\kappa$ and $\mathbb{C}_\gamma$ follow the classical dynamics, and 
quantum effects manifest only at higher orders in canonical operators. 
Gaussian states are characterized by their mean $\expval{\hat{\mathbf{r}}}$ and covariance matrix,
\begin{equation}
    \sigma_{ij} = \frac{1}{2}\expval{\left\{\hat{r}_i, \hat{r}_j\right\}} - \expval{\hat{r}_i}\expval{\hat{r}_j}
\end{equation}
where $\hat{\mathbf{r}} = (\hat{v}_+ \quad \hat{v}_-)^T$ and $\{\,\cdot\,,\,\cdot\,\}$ is the anticommutator. 
To highlight the quantum aspects of control we take $\expval{\hat{\mathbf{r}}} = 0$ so the system is classically controlled.
Because both $\mathbb{S}_\kappa$ and $\mathbb{C}_\gamma$ are Gaussian channels, Gaussian states remain Gaussian and we only need to track evolution of $\sigma$,
\begin{gather}
    \mathbb{S}_\kappa[\sigma] = \begin{pmatrix}
        e^{2\kappa}\sigma_{11} & \sigma_{12} \\ 
        \sigma_{12} & e^{-2\kappa}\sigma_{22}
    \end{pmatrix}\\
    \mathbb{C}_\gamma[\sigma] = e^{-2\gamma} \,\sigma + \frac{\hbar}{2}\left(1-e^{-2\gamma}\right) \,\mathbf{1}. \label{eq:control_channel}
\end{gather}
By design $\mathbb{S}_\kappa$ is a single-mode squeezing channel, and $\mathbb{C}_\gamma$ is a pure-loss attenuator~\cite{Serafini2021}.  
For $p\neq0$ the off-diagonal elements of $\sigma$ vanish for $t\to\infty$ due to control, so $\sigma_{11},\sigma_{22}\equiv \sigma_\pm$ alone determine steady state properties. 
These are the variances of canonical operators, so the constant term in $\mathbb{C}_\gamma$ manifests the quantum uncertainty relation between them, $\sigma_+\sigma_-\geq \hbar^2/4$, and the control point is $\sigma_\pm=\hbar/2$. 
We note now that \cref{eq:control_channel} is the basis for the noise inserted via the TWA control protocol mentioned earlier.
That noise is the classical representation of the uncertainty-limited broadening imposed by the quantum control channel.
The actions of $\mathbb{S}_\kappa$ and $\mathbb{C}_\gamma$ on $\sigma_\pm$ do not commute, complicating the analysis of trajectories of $\sigma_\pm$, however the system can be analyzed numerically together with analytical methods in applicable limits described later. 
For convenience we scale $\sigma_\pm$ by $\hbar$ in our further analysis, leaving them dimensionless.

\textit{Semiclassical analysis of quantum control transition}---As above we use $\bar\rho_{00}$ as the order parameter, with the controlled state signaled by $\bar{\rho}_{00}>0$. 
With our assumption $\expval{\hat{\mathbf{r}}}=0$, for a Gaussian trajectory $\rho_{00}(t) = \left[(\sigma_+(t)/+\tfrac{1}{2})(\sigma_-(t)+\tfrac{1}{2})\right]^{-1/2}$~\cite{Serafini2021}.
Since an arbitrary density matrix can be written as a linear combination of Gaussian density matrices, the order parameter $\bar{\rho}_{00}$ for a generic state is 
\begin{equation} \label{eq:rho00}
    \bar{\rho}_{00} = \int\dd \sigma_+\dd\sigma_- \frac{\mathcal{Q}(\sigma_+,\sigma_-)}{\sqrt{(\sigma_++\frac{1}{2})(\sigma_-+\frac{1}{2})}},
\end{equation}
where $\mathcal{Q}(\sigma_+,\sigma_-)$ is the steady-state probability density over Gaussian trajectories which obeys
\begin{equation} \label{eq:Frobenius}
    \mathcal{Q}(\sigma_+,\sigma_-) = (1-p)\mathcal{Q}(\mathbb{S}_\kappa^{-1}[\sigma_+,\sigma_-]) + p\,\mathcal{Q}(\mathbb{C}_\gamma^{-1}[\sigma_+,\sigma_-]).
\end{equation}
This is analogous to the Frobenius-Perron equation for the invariant density used in classical chaotic dynamics~\cite{Hasegawa1992}
~\footnote{In classical systems, this operator is ill defined for Hamiltonian dynamics, requiring the introduction of noise to regularize divergences~\cite{Supplement}. 
In contrast, quantum fluctuations resolve these divergences.}. 
We can compute \cref{eq:rho00} numerically.

\begin{figure}[t]
    \centering
    \includegraphics[width=0.87\columnwidth]{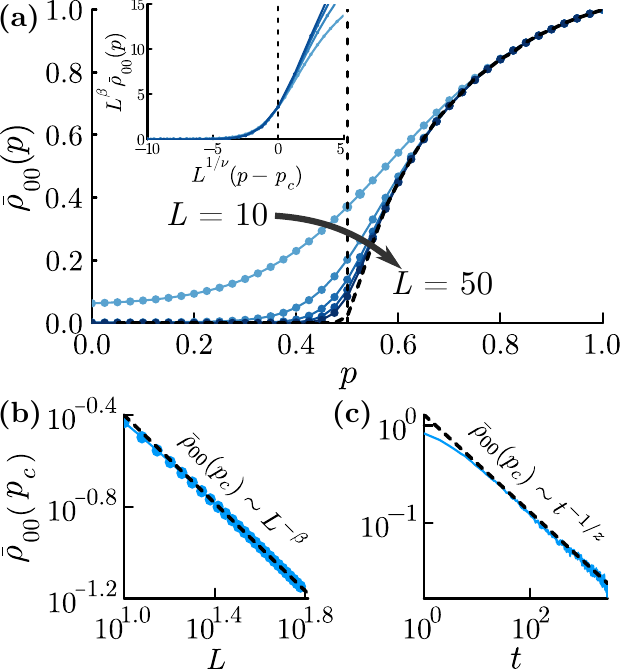}
    \caption{(a) Steady-state $\bar{\rho}_{00}(p)$ from the fixed point of $\mathbb{T}$ for $\kappa=\gamma=2\ln\!\left(\tfrac{1+\sqrt{5}}{2}\right)\approx0.42$, with cutoffs $2^{-L-1}<\sigma_\pm<2^{L-1}$ for $L=10,20,30,40,50$; dashed: late-time average over 5000 trajectories; inset: collapse near $p=p_c$ with $\nu=1$ and $\beta=0.96$. (b) $\bar{\rho}_{00}(L)$ at $p_c$ (same cutoffs); dashed: $\bar{\rho}_{00}\propto L^{-\beta}$ with $\beta=0.96$. (c) $\bar{\rho}_{00}(t)$ at $p_c$; dashed: $\bar{\rho}_{00}\propto t^{-1/z}$ with $z=2$.}
    \label{fig:rho00steady}
\end{figure}

In \cref{fig:rho00steady} we show results computed with the IHO corresponding to those presented for the cat map in \cref{fig:catmapfig}. 
The order parameter's dependence on $p$ shown in \cref{fig:rho00steady}(a) is computed both as an average over stochastic trajectories and as the fixed-point of the quantum channel after imposing cutoffs $\sigma_+ \in (1/2,2^{L-1})$ and $\sigma_- \in (2^{-L-1},1/2)$ \footnote{Due to normalization of $\sigma_\pm$ by $\hbar$, this cutoff defined by $L$ corresponds to taking $\hbar \sim 1/2^L$ similar to the cat map definition}.
The two methods agree for large $L$, with quick convergence for $p>p_c$ since values of $\sigma_\pm$ far from the control point are rarely explored.
Despite considering only classically controlled states $\expval{\hat{\mathbf{r}}}=0$ we see a clear control transition sharpening at $p_c$ for large $L$. 
Critical properties at $p_c$ are obtained from $L$ and $t$ dependence of $\bar{\rho}_{00}$, shown in Figs.~\ref{fig:rho00steady}(b),\ref{fig:rho00steady}(c), and by collapsing the $p$ dependence of $\bar\rho_{00}$ near $p=p_c$. 
We find $\beta \approx 1$ and $\nu\approx 1$, agreeing with the result of the cat map, and also dynamical exponent $z\approx2$, all characteristic of random walk universality. 

\textit{Fokker-Planck analysis}---Since $\sigma_- \in (0,1/2)$ at late times, $\bar{\rho}_{00}$ vanishing or remaining nonzero must result from $\sigma_+$ diverging for $p<p_c$ and remaining finite for $p>p_c$ in \cref{eq:rho00}.
The stochastic dynamics of $\sigma_+$ are a random multiplicative process with lower bound $\sigma_+>1/2$~\cite{Sornette1997}:
$\sigma_+(t+1)=\Lambda_t\, \sigma_+(t)$ where $\Lambda_t$ is drawn randomly from $(e^{2\kappa},e^{-2\gamma})$ with corresponding probabilities $(1-p,\, p)$, and $\sigma_+>1/2$ imposed by hand. 
In the limit of weak dynamics ($\kappa,\gamma$ small) and defining $y=\ln\sigma_+$, we derive a Fokker-Planck (FP) equation for $P(y,t)$, the distribution of values taken by $y$ during these dynamics,
\begin{equation} \label{eq:FP0}
    \pdv{P(y,t)}{t} = -\bar{v}\pdv{P(y,t)}{y} + \mathcal{D}\pdv[2]{P(y,t)}{y},
\end{equation}
with drift velocity $\bar{v} = 2\kappa\left(p_c-p\right)/p_c$ and diffusion constant $\mathcal{D} = 2\kappa^2 p(1-p)/p_c^2$.
This can be solved for both $P(y,t)$ far from the boundary $y^\mathrm{min}=-\ln2$ and for the nontrivial steady state $P(y)$ for $p>p_c$,
\begin{gather}
    P(y,t) \sim \frac{1}{\sqrt{2\pi \mathcal{D}t}}e^{-\frac{\left(y-\bar{v}t\right)^2}{2\mathcal{D}t}},\,\, y\gg y^\mathrm{min} \\
    P(y) = \frac{\xi}{2^\xi} e^{-\xi y} \,\,\Rightarrow\,\,     \mathcal{Q}_+(\sigma_+) = \frac{\xi}{2^\xi}  \sigma_+^{-1-\xi},\,\, p>p_c, \label{eq:powerlawQ}
\end{gather}
where $\xi = \abs{\bar{v}}/\mathcal{D}$. 
The distributions for $\sigma_+$ are obtained as $\mathcal{Q}_+(\sigma_+) = P(\ln\sigma_+)/\sigma_+$, and have log-normal and power-law forms, respectively~\footnote{The distribution $\mathcal{Q}_+(\sigma_+)$ is related to the full distribution $\mathcal{Q}(\sigma_+,\sigma_-)$ appearing in \cref{eq:rho00} and \cref{eq:Frobenius} by integrating over the allowed values of $\sigma_-$, giving a distribution for the values of $\sigma_+$ alone.}.
In Supplemental Material~\cite{Supplement} we present the details of the FP derivation and demonstrate this log-normal behavior of $\mathcal{Q}_+(\sigma_+,t)$.

With solutions for the distribution we can examine criticality of the transition. 
At $p=p_c$, $\bar{v}=0$ so the distribution has no overall drift and $y$ evolves diffusively away from the boundary, yielding a divergence in $y$ (and $\sigma_+$) at late times,
\begin{equation} \label{eq:zeqn}
    \bar{\rho}_{00}(t) \sim \int_{\frac{1}{2}}^\infty\!\!\!\dd\sigma_+\frac{\mathcal{Q}_+(\sigma_+,t)}{\sqrt{\sigma_+(t)}} \sim \int_{0}^{\infty}\!\!\!\dd y\frac{e^{-\frac{y^2}{2\mathcal{D} t}}}{\sqrt{\mathcal{D}t}} e^{-\frac{y}{2}} \sim t^{-1/2}. 
\end{equation}
If we also impose a maximum $\sigma_+^\mathrm{max} = 2^L$, giving $y^\mathrm{max}=L\ln2$, then for late times and large $L$ the hard wall boundaries combined with the diffusive nature of $P$ give $P(y)\sim1/L$ and we have
\begin{equation} \label{eq:betaeqn}
    \bar{\rho}_{00}(L) \sim \int_{\frac{1}{2}}^{2^L} \dd\sigma_+ \frac{\mathcal{Q}_+(\sigma_+)}{\sqrt{\sigma_+}}\sim\int_{0}^{L\ln2}\dd y \frac{e^{-\frac{y}{2}}}{L}\sim \frac{1}{L}.
\end{equation}
The Fokker-Planck analysis therefore recovers the critical behavior observed numerically in Figs.~\ref{fig:rho00steady}(b),\ref{fig:rho00steady}(c) with exponents $\beta=1$ and $z=2$. 

The multiplicative dynamics used to obtain \cref{eq:FP0} are approximate.
The map $\mathbb{C}_\gamma$ imposes the constraint $\sigma_+>1/2$ in a smooth way, while the analysis here imposed it crudely, with the difference largest for $\sigma_+\sim1/2$.
However, the universal results \cref{eq:zeqn,eq:betaeqn} depend only on the large-$\sigma_+$ properties of \cref{eq:FP0}---diffusive behavior away from the boundary and normalizability of distributions---and critical properties are insensitive to precise behavior near $\sigma_+\sim 1/2$.
This is demonstrated in Supplemental Material~\cite{Supplement} by derivation of the FP equation from exact $\mathbb{S}_\kappa$ and $\mathbb{C}_\gamma$. 

\textit{Exact quantum dynamics}---The FP analysis developed above is semiclassical, but exact quantum results can be obtained as well.
Since $\hat{\mathbf{1}}$ and $\hat{v}_+$ are eigenvectors of $\mathbb{T}$ with eigenvalues $1$ and $(1-p)e^{\kappa}+p\,e^{-\gamma}$, an inductive argument can be developed using these base cases to give an infinite set of eigenvectors
\begin{equation}
    \hat{Z}_n = (\hat{v}_+)^n + \sum_{k=1}^{\lfloor n/2\rfloor} \alpha_{n,k} \hat{Z}_{n-2k} = \sum_{k=0}^{\lfloor n/2\rfloor}\beta_{n,k}(\hat{v}_+)^{n-2k},
\end{equation}
with eigenvalues $\lambda_n = (1-p)e^{n\kappa}+p\,e^{-n\gamma}$, i.e., $\mathbb{T}[\hat{Z}_n] = \lambda_n \hat{Z}_n$, where the $\lfloor n/2\rfloor$ coefficients $\alpha_{n,k}$ or $\beta_{n,k}$ are found by solving a system of linear equations.
The full construction is detailed in Supplemental Material~\cite{Supplement}. 

When $\lambda_n<1$ the operator $\hat{Z}_n$ is controlled, $\langle\hat{Z}_n\rangle\to0$ as $t\to\infty$.
Since the eigenvalues of all other order-$n$ eigenoperators are less than one at $p=p_c$, control of $\hat{Z}_n$ diagnoses controllability for all operators of order $n$.
We thus obtain
\begin{equation} \label{eq:pstar}
    p^\ast_n = \frac{e^{n\kappa}-1}{e^{n\kappa}-e^{-n\gamma}} >p_c\quad \forall n>0
\end{equation}
as the control probability for all order-$n$ operators, with $\lim_{n\to0}p^\ast_n = p_c$.
Since $p^\ast_{n+1}>p^\ast_n>p_c$ for $n>0$, as $p$ increases above $p_c$ there is a sequence of $p$ values where operators of increasing order become finite.
In the End Matter, we relate the sequence $p_n^\ast$ to moments of the steady-state distribution obtained from our Fokker-Planck analysis becoming finite.

\textit{Conclusion}---We elucidate universal features of stochastic control transitions in quantum chaotic systems exemplified by the quantum cat map.
Agreement between exact dynamics, TWA, and the controlled IHO shows that these universal features are set by uncertainty-limited fluctuations and are largely insensitive to quantum interference.
This framework explains the emergence of random-walk universality and how quantum fluctuations dress the controlled phase.
Previous work~\cite{Iadecola2023,LeMaire2024,Allocca2024,Pan2024,Pan2025} demonstrated stochastic quantum control \emph{without} full canonical quantization and hence lacked a formal connection to quantum chaos.
In this work, uncertainty-limited localization and the operator-moment hierarchy provide a genuinely quantum signature; this is underscored by the fact that the small-$\hbar$ physics is simulable with a classical control map with quantum-limited noise \cite{Supplement}.
Nonetheless, due to the power-law nature of the steady-state \cref{eq:powerlawQ}, there is potential for interference phenomenon to be seen in the nature of higher moments beyond the analysis here; we leave that analysis to future work.

A direct platform for the IHO is a driven Kerr/parametric oscillator in cavity or circuit QED; above threshold the origin (vacuum) is an unstable fixed point and the dynamics form an effective phase-space double well while measurement-based feedback or autonomous reservoir engineering implements the attenuation control channel~\cite{Leghtas2015,Grimm2020,Sayrin2011,Vijay2012}. 
Collective-spin protocols also realize nonlinear dynamics and monitored entanglement regimes~\cite{Arias2020,Poggi2024}.
Fully chaotic maps with clear unstable orbits have cold atomic candidates, where analogous control protocol could be designed~\cite{Moore1995,Chaudhury2009}.

As noted above, the generic theory of the IHO cannot access the uncontrolled phase of specific maps.
Other quantum phenomena within that phase, such as qubit encoding, interference, and entanglement, can occur and are rich areas for future development.
Beyond these, a further question remains of how the \emph{many-body} problem behaves, where the phase space is extensive and Lyapunov exponents can undergo a transition themselves.

\begin{acknowledgments}
\emph{Acknowledgments}---We thank M.~Prasad, A.~Chakraborty, T.~Iadecola, M.~Kulkarni, and J.~H.~Pixley for discussions and collaboration on related work, and Y.~BenTov for discussions on relations to market dynamics.
J.~H.~W. is supported by the NSF CAREER Grant No.~DMR-2238895.
S.~G. is supported by NSF CAREER Grant No. DMR-1944967 and in part by NSF Grant No.~DMR-2315063.
This work was initiated and performed in part at the Aspen Center for Physics, which is supported by the National Science Foundation Grant No.~PHY-1607611. 
Portions of this research were conducted with high-performance computational resources provided by Louisiana State University~\cite{HPC}.
\end{acknowledgments}

\textit{Data Availability}---The data that support the findings of this article are openly available~\cite{data1,*data2}.

\bibliography{references}

\appendix

\onecolumngrid
\section{{\large End Matter}}
\twocolumngrid

Though our main focus has been on the control transition itself, we can examine generic properties of the controlled phase using both distributions obtained from the Fokker-Planck analysis and exact quantum channel results of the IHO. 

\begin{figure}[h]
    \centering
    \includegraphics[width=0.8\columnwidth]{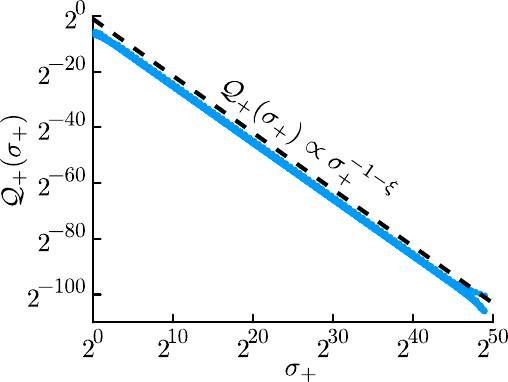}
    \caption{Comparison of the power-law form of the distribution $\mathcal{Q}_+(\sigma_+)$ acquired from the FP analysis with the distribution of $\sigma_+$ computed numerically from the fixed point of $\mathbb{T}$ with $\kappa = \gamma = 0.2$ and cutoff $L=50$ for $p=0.6$.}
    \label{fig:powerlaw}
\end{figure}

In \cref{fig:powerlaw}, we compare the power-law form of $\mathcal{Q}_+(\sigma_+)$ in \cref{eq:powerlawQ} to the distribution computed numerically for $p>p_c$.
The lower boundary of the FP dynamics is vital for the existence of a late-time steady state in the controlled phase---in \cref{eq:FP0} $\bar{v}<0$ for $p>p_c$, so the distribution drifts to small $y$ and is held up by the barrier at $y^\mathrm{min}$. 
These two opposing forces balance to yield a stable distribution. 

We can further compute moments of this distribution
\begin{equation}
    \overline{\sigma_+^n} = \int\dd\sigma_+ \sigma_+^n\mathcal{Q}_+(\sigma_+) = \frac{2^{-\xi}\xi}{\xi-n},
\end{equation}
and find that the $n$th moment of $\sigma_+$ is finite only when $\xi>n$.
Setting $\xi = n$ gives a sequence of control rates at which these moments become finite in the FP framework, 
\begin{equation} \label{eq:pFP}
    p^\mathrm{FP}_{2n} = p_c + (1-p_c)n\kappa + O(\kappa^2) > p_c,
\end{equation}
where we have expanded the result in small $\kappa$ since this was necessary to obtain \cref{eq:FP0} and $\mathcal{Q}_+(\sigma_+)$. 
The $n$th moment of $\mathcal{Q}_+$ is related to quantum operators of order $2n$ since $\sigma_+$ is quadratic.

This sequence of moment control rates are directly related to the sequence of control rates $p^\ast$ in \cref{eq:pstar} for which quantum operators of order $n$ become controlled: $p^\ast_{2n} = p^\mathrm{FP}_{2n}$ to first order in $\kappa$. 
This confirms that the FP analysis captures key features of the exact quantum dynamics in the appropriate limits, and equivalently demonstrates that the operators $\hat{Z}_{n}$ generalize the moment-control picture of the semiclassical analysis to a quantum context.

\vfill




\end{document}


\title{Supplement to Universality of stochastic control of quantum chaos with measurement and feedback}

\author{Andrew A. Allocca}
\affiliation{Department of Physics and Astronomy, Louisiana State University, Baton Rouge, Louisiana 70803, USA}
\affiliation{Center for Computation and Technology, Louisiana State University, Baton Rouge, Louisiana 70803, USA}

\author{Devesh K. Verma}
\affiliation{Department of Physics and Astronomy, Louisiana State University, Baton Rouge, Louisiana 70803, USA}

\author{Sriram Ganeshan}
\affiliation{Department of Physics, City College, City University of New York, New York, New York 10031, USA}
\affiliation{CUNY Graduate Center, New York, New York 10031, USA}

\author{Justin H. Wilson}
\affiliation{Department of Physics and Astronomy, Louisiana State University, Baton Rouge, Louisiana 70803, USA}
\affiliation{Center for Computation and Technology, Louisiana State University, Baton Rouge, Louisiana 70803, USA}

\maketitle




\section{The Quantum Arnold Cat Map details}

\subsection{Alternative derivation}

To derive the cat map, we take the Heisenberg evolution perspective and attempt to find a unitary matrix such that
\[
U^\dagger \hat x U = 2\hat x + \hat p, \quad U^\dagger \hat p U = \hat x + \hat p.
\]
Further, note that $[\hat x, \hat p] = i \hbar$. We will use $\hbar$ to compactify us onto the torus.

Two operations will be helpful:
\[
\begin{aligned}
e^{i \alpha \hat p^2} \hat x e^{-i \alpha \hat p^2} & = \hat x + i \alpha [\hat p^2, \hat x] = \hat x + 2\alpha\hbar  \hat p \\
e^{-i \beta \hat x^2} \hat p e^{i \beta \hat x^2} & = \hat p - i \beta[\hat x^2, \hat p] = \hat p + 2\beta \hbar  \hat x 
\end{aligned}
\]
If we let $\alpha = \frac1{2\hbar}$ and $\beta = \frac{1}{2\hbar}$, then notice that to get the cat map for Heisenberg evolution, we must have
\begin{equation}
   U_\mathrm{cat} = e^{-i \hat p^2 / (2\hbar)} e^{i \hat x^2 / (2\hbar)}
\end{equation}

Equivalently, using Berry's method of quantizing linear maps~\cite{Hannay1980}, we generate a quantum version of Arnold's cat map~\cite{Arnold1968}, which is a well known chaotic map.
The classical cat map is defined to act on a two-dimensional parameter space as
\begin{equation}
\label{eq:catmapSUPP}
  \begin{pmatrix} x' \\ p' \end{pmatrix}=
  \begin{pmatrix} 2 & 1 \\ 1 & 1 \end{pmatrix} 
  \begin{pmatrix} x \\ p \end{pmatrix} \mod 1.
\end{equation}

The quantized cat map propagator acts on states $\ket{Q}$ for $Q = 0,\ldots, N-1$ (and $x = Q/N$ is the point in the phase space) is given by:
\begin{equation}
    \expval{Q_1 | U | Q_2}=\sqrt{\frac{i}{N}}\exp\Big(\frac{i \pi}{N}(Q_1^2+(Q_2 - Q_1)^2 \Big)
\end{equation}
where $N$ must be a positive even integer.
We can map this to phase space $\ket{Q} = \frac1N \sum_P e^{2\pi i Q P / N} \ket{P}$, where now $P=0,\ldots N-1$, and if we want to impose $p = P/N$, this necessarily means $e^{2\pi i Q P / N} \rightarrow e^{2\pi N i x p} = e^{i x p / \hbar}$, which works if we let
$$
\hbar = \frac1{2\pi N}.
$$
Therefore, finite-size in these compact discrete maps is intimately tied to its classicality.

Near the (unstable) fixed point $(0,0)$, the cat map dynamics can be approximated by an inverted harmonic oscillator, and we can derive exactly how by finding the eigenvalues and eigenvectors of the map in \cref{eq:catmapSUPP} in terms of $\varphi=\frac{1+\sqrt{5}}2$, for which we have $\{ \varphi^2, \varphi^{-2}\}$ as the eigenvalues and $\{(\varphi,1)^T, (-\varphi^{-1},1)^T\}$ as the (unnormalized) eigenvectors.

This model, therefore has $\kappa = 2\ln \varphi$.

\subsection{Control operation}

In order to control this system onto the unstable fixed point at the origin, we need a space in which to build the $a$ and $a^\dagger$ operators used in the Kraus operators in the main text. 

These operators should annihilate the vacuum state $\ket{0}$ and approximately create harmonic oscillator numbers states $\ket{n}$ when $n \ll N$.
Respecting periodic boundary conditions, we construct these states from the following ``control Hamiltonian'' (we do not use this for dynamics, but purely to find a complete set of states to define these operators):
$$
\hat H_{\mathrm{ctrl}} = 2 - \cos(2\pi \hat x) - \cos(2\pi \hat p).
$$
First, a note about the energy structure: This Hamiltonian classically has a separatrix at $E=2$, i.e., the central diamond centered around (0,0) and defined by $|x-\frac12|+|y-\frac12|=\frac12$ (this defines the region $\Omega$ mentioned in the main text). 
This naturally leads us to exclude things outside of the diamond containing $(0,0)$.
Quantum mechanically, it means only using states with $0<E<2$ for control; this will make no difference to the in-text analysis of the controlled phase.

With $\hat H_{\mathrm{ctrl}}$, we can define the ladder operator $a$ via the eigenstates $\hat H_{\mathrm{ctrl}} \ket{n} = E_n \ket{n}$, so that
$$
a \ket{n} = \sqrt{n} \ket{n-1}.
$$
However, since $a$ connects states with different energies, the global phase chosen for those states changes the operator $a$ itself.
To fix this phase, we first choose $\ket{n}$ to be all real valued, and we then fix the sign of $a$ by imposing the standard harmonic oscillator signs for wave functions $\braket{x}{n} = \psi_{n}(x)$,
$$
\sgn(\psi_{2n}(0)) = (-1)^{n}, \quad \sgn(\psi'_{2n+1}(0)) = (-1)^n.
$$
This fixes the operator $a$ within the diamond defined by the $0<E<2$ states (inside the classical separatrix).

With this definition for $a$, the Kraus operators need no modification except to be truncated to a smaller space $n=0,\ldots N-1$.
The control procedure does get slightly modified though: We first measure if the quantum state is in $0<E<2$, and if not, we do nothing.
If it is, we then choose which Kraus operator $K_m$ to apply based on its Born probability and proceed as normal.
The quantum simulations are then done on \emph{trajectories} to maintain state purity and avoid simulating full density matrices.
Therefore, when control is done, we choose which Kraus operator $K_m$ to apply based on its Born probability
\begin{equation}
    P^{\mathrm{Born}}_{m}=\bra{\psi}\hat{K}^{\dagger}_m\hat{K}_m\ket{\psi}.
\end{equation}
where, $\ket{\psi}$ is the system wavefunction and $0\leq m \leq N-1$.

We choose $p$ to be the measurement rate, and perform the unitary cat map evolution at a rate of $1-p$.
Performing this stochastic evolution, we can average over trajectories  to obtain quantities like the order parameter $\bar{\rho}_{00}(t) = \overline{\lvert\expval{0|\psi_t}\rvert^2}$, where $\overline{(\cdots)}$ represents averaging over trajectories.

\textbf{Matching the classical control}---To match classical control, we need only to condition it on the region
$$
 \mathbf r \mapsto \mathbf r_0 + e^{-\gamma} (\mathbf r - \mathbf r_0), \quad |x-\tfrac12|+|p-\tfrac12| > \tfrac12.
$$
for $\mathbf r = (x,p)$ and $\mathbf r_0$ is defined such that
$$
\mathbf r_0 = \begin{cases}
 (0,0), & x + p < 1/2, \\
 (1,0), & p - x > 1/2, \\
 (0,1), & x - p > 1/2, \\
 (1,1), & x + p > 3/2,
\end{cases}
$$
in order to satisfy the periodic boundary conditions appropriately.
While this is not an exact mapping to the Kraus evolution described above, it is a very good approximation as is demonstrated via linear approximation in the main text Eq.~(9).

\subsection{Truncated Wigner approximation}

The semiclassical simulations use the truncated Wigner approximation (TWA)~\cite{Polkovnikov2010} with phase space lifted to $\mathbb R^2$, valid when trajectories remain localized near the fixed point (torus winding and discrete Wigner effects are negligible).
With $[\hat x,\hat p]=i\hbar$ and $\hbar=1/(2\pi N)$, quantum states near the origin are approximated by smooth Wigner distributions $W(\mathbf r)$ for $\mathbf r=(x,p)$.
The target vacuum has Wigner function
\begin{equation}
    W_0(\mathbf r) = \frac{1}{\pi\hbar}e^{-\mathbf r^2/\hbar},
\end{equation}
where we use local coordinates centered at the fixed point.

\textbf{Unitary step.}---The cat-map propagator $\mathbf r\mapsto M\mathbf r \mod 1$ with $M=\begin{pmatrix}2&1\\1&1\end{pmatrix}$ from \cref{eq:catmapSUPP} acts by pushforward,
\begin{equation}
    W_{t+1}(\mathbf r) = W_t(M^{-1}\mathbf r).
\end{equation}

\textbf{Control step.}---Near the fixed point the measurement-feedback channel acts as a pure-loss attenuator on canonical variables (see \cref{sec:classicalnoise} for the classical noisy analog).
The affine update is
\begin{equation}
    \mathbf r \mapsto e^{-\gamma}\mathbf r + \boldsymbol{\eta},\qquad 
    \boldsymbol{\eta}\sim \mathcal N\!\left(\mathbf 0,\; \tfrac{\hbar}{2}\left(1-e^{-2\gamma}\right)\mathbf 1\right),
\end{equation}
with the restriction to the region $|x-\tfrac12|+|p-\tfrac12| > \tfrac12$ as discussed above.

\textbf{Stochastic trajectories.}---We sample $\mathbf r(0)\sim W_0$ and stochastically alternate the two maps: with probability $(1-p)$ apply $\mathbf r\mapsto M\mathbf r \mod 1$, and with probability $p$ apply control.
Observables are computed by averaging over the ensemble.

\textbf{Fidelity.}---The order parameter $\rho_{00}(t)=\mel{0}{\rho(t)}{0}$ is estimated via the Wigner overlap,
\begin{equation}
    \rho_{00}(t)\approx 2\pi\hbar\int d^2\mathbf r\; W(\mathbf r,t)\,W_0(\mathbf r) \approx \frac{2\pi\hbar}{N_\mathrm{samp}}\sum_{i=1}^{N_\mathrm{samp}} W_0\!\left(\mathbf r_i(t)\right),
\end{equation}
where $\mathbf r_i(t)$ are the $N_\mathrm{samp}$ sampled phase-space points at time $t$ and $W_0 = \frac{1}{\pi\hbar}e^{-\mathbf r^2/\hbar}$, the Wigner function of the target vacuum.
This produced the semiclassical results in Fig.~1 in the main text.

\subsection{Comparison of quantum cat map simulations and TWA}
\begin{figure}[h]
    \centering
    \includegraphics[width=\linewidth]{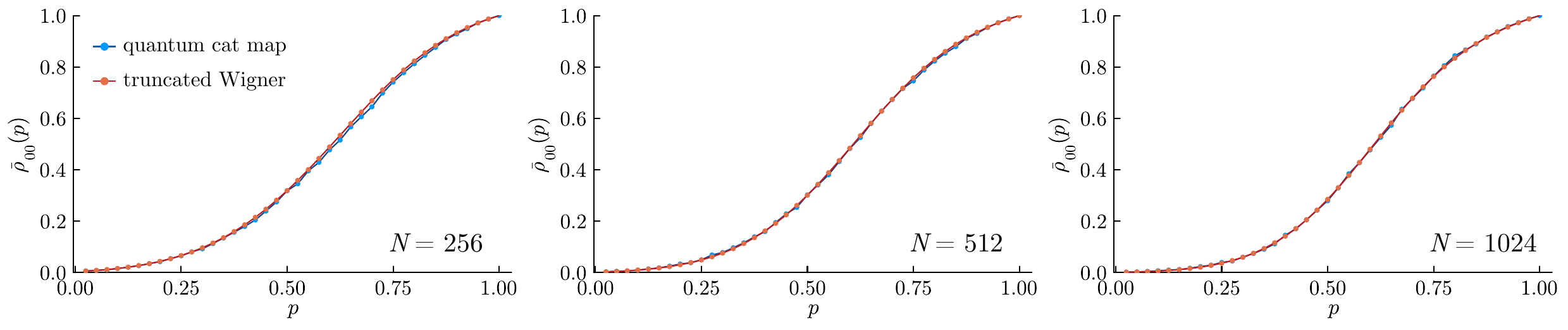}
    \caption{Comparison between $\bar{\rho}_{00}(p)$ computed from simulation of the exact dynamics of the quantum cat map (blue) and the truncated Wigner approximation (red), for three different system sizes $N=256,512,1024$. (Truncated Wigner has quantum noise due to $\hbar = \frac1{2\pi N}$, see text.)}
    \label{fig:comparison}
\end{figure}

For large enough system sizes the order parameter $\bar{\rho}_{00}$ obtained from direct simulation of the quantum cat map and from the TWA of these dynamics are in excellent agreement, demonstrated in \cref{fig:comparison}.
This justifies using the TWA to extend the results for the quantum cat map to much larger system sizes than can be exactly simulated; in the main text we consider the quantum cat map results up to $N=1024$ ($L=\log_2N=10$) and then extend to larger $N$ with the TWA. 

\section{Fokker-Planck derivation}

When the system is initialized in a state such that $\expval{\hat{x}}=\expval{\hat{p}}=0$, the late-time value of $\bar{\rho}_{00}=\overline{\mel{0}{\rho}{0}}$, which indicates whether the system is controlled, is determined by just a single element of the covariance matrix, $\sigma_+ = \expval{(\hat{v}_+)^2}$. 
With the known form of the channels $\mathbb{S}_\kappa$ and $\mathbb{C}_\gamma$, the evolution of $\sigma_+$ can be expressed as a modified multiplicative process:
\begin{equation}
    \sigma_+(t+\delta t) = e^{f(\sigma_+,v)}e^{v\,\delta t}\,\sigma_+(t),
\end{equation}
where $v$ is drawn randomly from the distribution
\begin{equation}
    \pi(v) = (1-p)\,\delta(v-2\Omega)+p\,\delta(v+2\Gamma)
\end{equation}
and
\begin{equation}
    f(\sigma_+,v) = \log\left(1+h(v)\frac{e^{-v\,\delta t}-1}{2 \,\sigma_+}\right),
\end{equation}
with $h(2\Omega)=0$ and $h(-2\Gamma)=1$; the form of $h(v)$ is not yet determined beyond the values it takes at these two points.
The factor $e^{f(\sigma_+,v)}$ comes from the term in the control evolution imposing the lower bound on $\sigma_+$ from the uncertainty relation. 
The parameters we use for the strength of the quantum channels are $\kappa = \Omega\delta t$ and $\gamma = \Gamma\delta t$.
The time step length $\delta t$ is superfluous in principle; only the strengths of the maps $\kappa$ and $\gamma$ are relevant for the dynamics.
Including this $\delta t$, however, allows us to more easily take a continuous time limit later in the derivation while avoiding pathological behavior of the yet-undetermined function $h(v)$. 

Putting $\sigma_+ = \frac{1}{2}e^{x}$ we can rewrite the dynamics in terms of a modified random walk of the exponential variable $x$~\cite{Sornette1997}.
The factor of $\frac{1}{2}$ is chosen so that the control point $\sigma_+ = \frac{1}{2}$ corresponds to $x=0$.
The time evolution of $x$ is then
\begin{equation}
    x(t + \delta t) = T(x(t),v) = x(t) + v \,\delta t+ \tilde{f}(x(t),v),
\end{equation}
where $\tilde{f}(x,v) = f(\tfrac{1}{2}e^x,v)$. 
The preimage of $x(t)$ for a given $v$ is
\begin{equation}
    T^{-1}(x,v) = x - v\,\delta t + \tilde{g}\left(x,v\right) \quad \text{with}\quad \tilde{g}(x,v) = \log\left[1-h(v)\left(1-e^{v\,\delta t}\right)e^{-x}\right].
\end{equation}
Letting $\mathcal{P}(x,t)$ be the distribution of values $x$ takes after many time steps, we have
\begin{align}
    \mathcal{P}&(x, t+\delta t) = \int_{-\infty}^\infty\dd v\,\pi(v)\,\mathcal{P}\left(T^{-1}(x,v),t\right) = \int_{-\infty}^\infty\dd v\,\pi(v)\,\mathcal{P}\left(x-v\,\delta t+\tilde{g}(x,v),t\right)\\
    &= \int_{-\infty}^\infty\dd (\delta v)\,\pi(\bar{v}+\delta v)\,\mathcal{P}\left(x-\bar{v}\,\delta t-\delta v\,\delta t+\tilde{g}(x,\bar{v}+\delta v),t\right) \\
    &\approx \mathcal{P}(T^{-1}(x,\bar{v}),t) + \delta t^2\frac{\expval{\delta v^2}}{2} \Bigg[\frac{1}{\delta t^2}\eval{\pdv[2]{\tilde{g}(x,v)}{v}}_{v=\bar{v}} \pdv{\mathcal{P}(y,t)}{y} + \left(1-\frac{1}{\delta t}\eval{\pdv{\tilde{g}(x,v)}{v}}_{v=\bar{v}}\right)^2 \pdv[2]{\mathcal{P}(y,t)}{y}\Bigg]_{y=T^{-1}(x,\bar{v})}, \label{eq:discreteFP}
\end{align}
where in the second line we put $v=\bar{v}+\delta v$ with $\bar{v} = \int\dd v\,v\,\pi(v) = 2\Omega-2p(\Gamma+\Omega)$ the average value of $v$, so that $\int\dd v\,\delta v\,\pi(v) = 0$.
In the final line we expand in $\delta v$, which involves expanding both the distribution $P$ and the function $\tilde{g}$.
This assumes that $\pi(v)$ is relatively sharply peaked around $\bar{v}$. 
We also pull a factor of $\delta t^2$ from the entire bracketed expression. 
Using one of these factors of $\delta t$, the expectation value of $\delta v^2$ can be manipulated to give a diffusion constant,
\begin{align}
    \mathcal{D} &\equiv \delta t\frac{\expval{\delta v^2}}{2} = \frac{\delta t}{2}\int_{-\infty}^\infty\dd v \,\pi(v)\,(v-\bar{v})^2 \\
    &= \frac{\delta t}{2}\left[4p^2(1-p)(\Gamma+\Omega)^2+4p(1-p)^2(\Gamma+\Omega)^2\right] = 2p(1-p)(\Gamma+\Omega)^2\delta t.
\end{align}
If we define a step size $\ell = v\delta t$ then we see that the diffusion constant is the mean-square step size per time step. 
It grows with the difference between the two possible values of $v$---if they are very different, then trajectories will diverge quickly, i.e. fast diffusion. 

Now consider a limit such that our dynamics are well-approximated as continuous in time. 
We can accomplish this by considering $\delta t$ small with fixed $\Omega,\Gamma$, but we should think of this as the limit of weak maps, i.e. small $\kappa,\gamma$. 
Expanding in $\delta t$ helps us avoid issues with the yet-undetermined function $h(v)$, but in the end the value of $\delta t$ is set to 1. 

Returning to \cref{eq:discreteFP}, we bring $\mathcal{P}(T^{-1}(x,\bar{v}),t)$ to the left-hand side and divide both sides by $\delta t$ giving
\begin{equation}
    \frac{1}{\delta t}\left[\mathcal{P}(x,t+\delta t)-\mathcal{P}(T^{-1}(x,\bar{v}),t)\right] 
    \underset{\delta t \text{ small}}{\longrightarrow} \pdv{\mathcal{P}(x,t)}{t} + \bar{v}\left(1-h(\bar{v})e^{-x}\right)\pdv{\mathcal{P}(x,t)}{x},
\end{equation}
on the left-hand side, where we have expanded $T^{-1}(x,\bar{v}) = x - \bar{v}\,\delta t + \tilde{g}(x,\bar{v})$ using $\tilde{g}(x,v)\approx\delta t\,v\,h(v)\,e^{-x}+\delta t^2\frac{v^2}{2}(1-h(v)e^{-x})h(v)e^{-x}$.
This could alternatively be written as a total time derivative with $x(t)$ evolving ballistically with velocity $\bar{v}$.

The right-hand side of \cref{eq:discreteFP} has factors related to the derivatives of $\tilde{g}(x,v)$,
\begin{gather}
    \frac{1}{\delta t}\eval{\pdv{\tilde{g}(x,v)}{v}}_{v=\bar{v}} = \eval{\pdv{\left(v\,h(v)\right)}{v}}_{v=\bar{v}}\,e^{-x} + O(\delta t) \\
    \frac{1}{\delta t^2}\eval{\pdv[2]{\tilde{g}(x,v)}{v}}_{v=\bar{v}} = \frac{1}{\delta t}\eval{\pdv[2]{\left(v\,h(v)\right)}{v}}_{v=\bar{v}}e^{-x} +
    \frac{1}{2}\eval{\pdv[2]{\left[v^2(1-h(v)e^{-x})h(v)\right]}{v}}_{v=\bar{v}}e^{-x} + O(\delta t). \label{eq:d2gdv2}
\end{gather}
The $1/\delta t$ in the first term of \cref{eq:d2gdv2} presents a problem.
Though we need to expand the derivatives of $\mathcal{P}$ in powers of $\delta t$ since they are evaluated at $T^{-1}(x,\bar{v})$ and not $x$, when the $1/\delta t$ term of \cref{eq:d2gdv2} multiplies $\partial \mathcal{P}(x,t)/\partial x$ there is no factor of $\delta t$ to cancel it, leading to a divergence for small $\delta t$ unless we impose a condition to cancel it in general,
\begin{equation}
    \pdv[2]{\left(v\,h(v)\right)}{v} = 0 \quad \Rightarrow \quad h(v) = \frac{\Gamma}{v}\frac{v-2\Omega}{\Gamma+\Omega}.
\end{equation}
Since $h(v)$ was not determined when introduced besides its value at $v=2\Omega$ and $v=-2\Gamma$ we are free to use this form to satisfy this additional constraint and ensure a well-formed Fokker-Planck equation in the continuous time limit. 
As a result of this condition, no additional terms involving additional derivatives of $\mathcal{P}$ will appear.
The needed derivatives are then
\begin{gather}
    \eval{\pdv{(v\,h(v))}{v}}_{v=\bar{v}} = \frac{\Gamma}{\Gamma+\Omega} = \frac{\gamma}{\kappa+\gamma} = 1-p_c \equiv \lambda \\
    \frac{1}{2}\eval{\pdv[2]{\left[v^2(1-h(v)e^{-x})h(v)\right]}{v}}_{v=\bar{v}}e^{-x} = \left(1-\frac{\Gamma}{\Gamma+\Omega}e^{-x}\right)\frac{\Gamma}{\Gamma+\Omega}e^{-x} = \left(1-\lambda e^{-x}\right)\lambda e^{-x}
\end{gather}

We therefore have the Fokker-Planck equation
\begin{equation} \label{eq:FokkerPlanck1}
    \pdv{\mathcal{P}(x,t)}{t} = -v(x)\pdv{\mathcal{P}(x,t)}{x} + \mathcal{D}\,r(x) \pdv{}{x}\left[r(x)\pdv{\mathcal{P}(x,t)}{x}\right],
\end{equation}
with
\begin{gather}
    r(x) = 1-\lambda e^{-x}\\
    v(x) = \bar{v}\left(1-\,h(\bar{v})e^{-x}\right) = \left(\bar{v}+ 2\kappa \frac{\lambda}{e^x-\lambda}\right)r(x).
\end{gather}
Notice that we have a factor of $r(x)$ with every spatial derivative, suggesting that we can change coordinates to make \cref{eq:FokkerPlanck1} take a cleaner form.
Putting
\begin{equation}
    z \equiv \log\left(e^x-\lambda\right) \,\Rightarrow \pdv{}{z} = r(x)\pdv{}{x},
\end{equation}
and defining $P(z,t) = \mathcal{P}\left(x(z),t\right)$ we obtain
\begin{align} \label{eq:FokkerPlanck2}
    \pdv{P(z,t)}{t} &= -(\bar{v}+2\gamma\,p_c e^{-z})\pdv{P(z,t)}{z} + \mathcal{D}\pdv[2]{P(z,t)}{z} \\
    &= -2\gamma\, p_c e^{-z} P(z,t)-\pdv{\left[\left(\bar{v}+2\gamma\, p_c e^{-z}\right)P(z,t)\right]}{z} + \mathcal{D}\pdv[2]{P(z,t)}{z}.
\end{align}
As a result, the original element of the covariance matrix is $\sigma_+ = \frac{1}{2}e^x = \frac{1}{2}(e^z-p_c+1)$, with the control point $\sigma_+=\frac{1}{2}$ now equating to $z = \log(p_c)$.

\begin{figure}[t]
    \centering
    \includegraphics[width=0.7\textwidth]{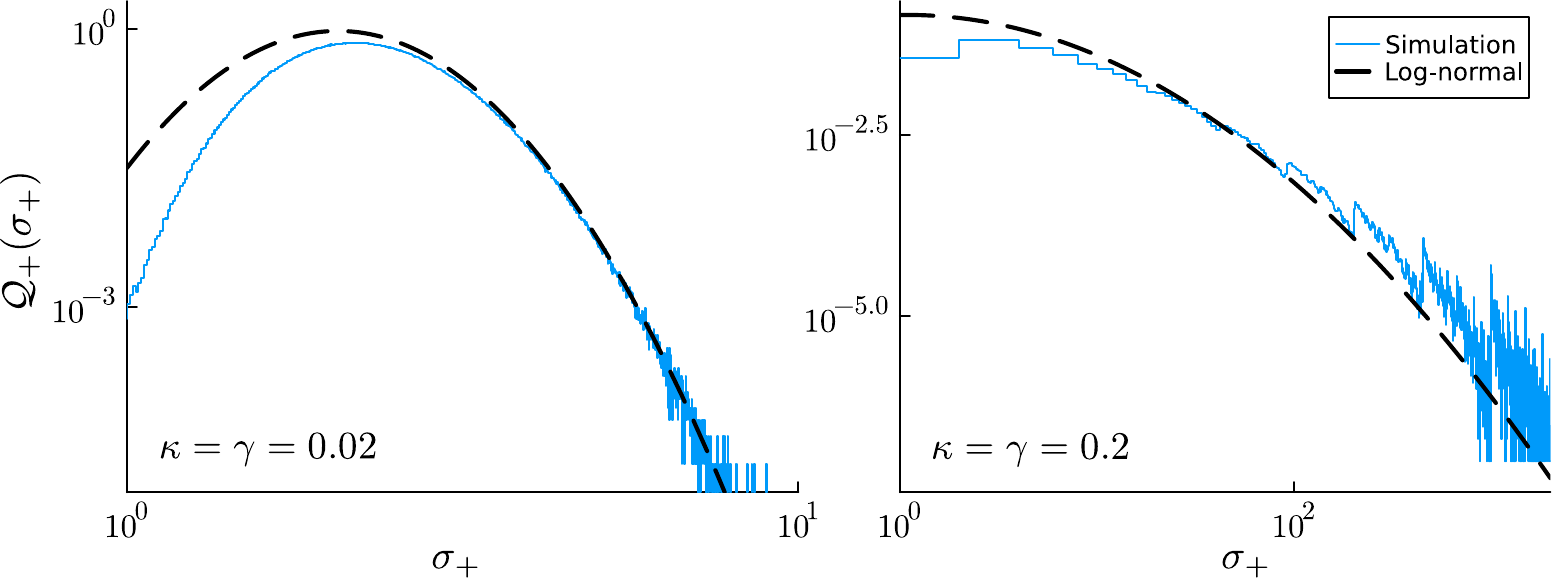}
    \caption{The early-time distribution of values taken by $\sigma_+$ in numerical simulations with $p=0.4<p_c$ compared to the predicted log-normal form of $\mathcal{Q}_\sigma(\sigma_+)$, shown in a log-log plot.
    The simulations initialize $\sigma_+(0)=1/2$ and show the values of $\sigma_+(t_f)$ across $5 \times 10^6$ stochastic trajectories.
    For $\kappa=\gamma=0.02$ we take $t_f=25$, and for $\kappa=\gamma=0.2$ we take $t_f=100$.}
    \label{fig:lognormal}
\end{figure}

The Fokker-Planck equation obtained here has followed from the exact stochastic dynamics on $\sigma_+$, and we see that the equation analyzed in the main text, 
\begin{equation} 
    \pdv{P(z,t)}{t} = -\bar{v}\pdv{P(z,t)}{z} + \mathcal{D}\pdv[2]{P(z,t)}{z},
\end{equation}
along with constraint $z>-\ln(2)$, agrees with this exact equation in the large-$z$ limit. 
Accordingly, the solutions for the distribution $P(z,t)$ and therefore $\mathcal{Q}_+(\sigma_+)$ far from this boundary are also valid, power-law at $p=p_c$ and log-normal for early times and $p\neq p_c$.
The power-law critical behavior is discussed in the main text, and in \cref{fig:lognormal} we also show agreement with the log-normal distribution in the appropriate limits.

\section{Eigenvectors of the Stochastic Evolution}

The stochastic dynamics we consider generates time evolution in the Heisenberg picture as
\begin{equation}
    \mathbb{T}[\hat{\mathcal{O}}(t+1)] = (1-p) \mathbb{S}_\kappa[\hat{\mathcal{O}}(t)] + p\, \mathbb{C}_\gamma[\hat{\mathcal{O}}(t)],
\end{equation}
where $\hat{\mathcal{O}}$ is a generic operator, and $\mathbb{S}_\kappa$ and $\mathbb{C}_\gamma$ are the squeezing and control channels. 
While all the eigenvectors of the individual channels $\mathbb{S}_\kappa$ and $\mathbb{C}_\gamma$ are simple to calculate, eigenvectors for the full channel $\mathbb{T}$ are not as straightforward to obtain.
Here, we now construct an infinite set of these eigenvectors that will allow us to draw important conclusions about the controlled phase of this model.

We first note the useful identities
\begin{gather}
    (\hat{v}_+)^k = \sum_{l=0}^{\floor{k/2}} \underbrace{\frac{(2l-1)!!}{2^l} \binom{k}{2l}}_{\equiv \mathcal{A}_{k,k-2l}} \,:(\hat{v}_+)^{k-2l}: \,\, =\,\, :(\hat{v}_+)^k: + \sum_{l=1}^{\floor{k/2}}\mathcal{A}_{k,k-2l}\,:(\hat{v}_+)^{k-2l}:\label{eq:id1}\\
    :(\hat{v}_+)^k:\,\, = \sum_{l=0}^{\floor{k/2}} \underbrace{(-1)^l \frac{(2l-1)!!}{2^l} \binom{k}{2l}}_{\equiv (\mathcal{A}^{-1})_{k,k-2l}} (\hat{v}_+)^{k-2l},\label{eq:id2}
\end{gather}
allowing us to transform between $(\hat{v}_+)^k$ and $:(\hat{v}_+)^k:$, where
\begin{equation}
    :(\hat{v}_+)^k:\, = \left(\frac{e^{-i\frac{\pi}{4}}}{\sqrt{2}}\right)^k\sum_{j=0}^k\binom{k}{j}(i\hat{a}^\dagger)^j\hat{a}^{k-j}
\end{equation}
has normally-ordered ladder operators, so that
\begin{equation}
    \mathbb{C}_\gamma[:(\hat{v}_+)^k:] = e^{-k\gamma} :(\hat{v}_+)^k:. 
\end{equation}
\cref{eq:id1} is simple to derive, and we obtain \cref{eq:id2} by noting that $\mathcal{A}_{k,k'}$ are the components of a triangular matrix---we have only nonzero elements for $k>k'$. 
The properties of triangular matrices then allow direct computation of $\mathcal{A}^{-1}$. 

Two low-order eigenvectors of $\mathbb{T}$ are straightforward: 
\begin{align}
    \mathbb{S}_\kappa[\hat{1}] = \mathbb{C}_\gamma[\hat{1}] = \hat{1}  \quad&\Rightarrow\quad \mathbb{T}[\hat{1}] = \hat{1}\\
    \mathbb{S}_\kappa[\hat{v}_+] = e^{\kappa}\hat{v}_+, \quad\mathbb{C}_\gamma[\hat{v}_+] = e^{-\gamma}\,\hat{v}_+ \quad&\Rightarrow\quad \mathbb{T}[\hat{v}_+] = \left[(1-p)e^{\kappa} + p\,e^{-\gamma}\right]\hat{v}_+.
\end{align}
so we can define the corresponding eigenvalues $\lambda_0 = 1$ and $\lambda_1(p) = (1-p)e^{\kappa}+p\,e^{-\gamma}$. 
We now build eigenvectors of $\mathbb{T}$ inductively using these as base cases.

Assume that for some fixed $n$ we have a set of operators $\{\hat{Z}_k\}$ for $k=0,2,\dots,n-2$ ($n$ even) or $k=1,3,\dots,n-2$ ($n$ odd), with $\hat{Z}_0\equiv\hat{1}$, $\hat{Z}_1\equiv \hat{v}_+$, and
\begin{equation}
    \mathbb{T}[\hat{Z}_k] = \lambda_k(p) \hat{Z}_k,
\end{equation}
with known eigenvalues $\lambda_k(p)$.
Additionally, assume that we know how to write any of these $\hat{Z}_k$ as a sum over the operators in this set with smaller index as
\begin{equation}
    \hat{Z}_k = (\hat{v}_+)^k + \sum_{l=1}^{\floor{k/2}} \alpha_{k,k-2l} \hat{Z}_{k-2l} = \sum_{l=0}^{\floor{k/2}} \beta_{k,k-2l} (\hat{v}_+)^{k-2l}, \label{eq:Zeq}
\end{equation}
where all coefficients $\alpha_{k,k-2l}$ are known and the coefficients $\beta_{k,k-2l}$ can be written in terms of the $\alpha$s by expanding all $\hat{Z}_{k-2l}$ in this same way and collecting powers of $\hat{v}_+$---the first few are $\beta_{k,k}=1$, $\beta_{k,k-2}=\alpha_{k,k-2}$, and $\beta_{k,k-4}=\alpha_{k,k-4}+\alpha_{k,k-2}\alpha_{k-2,k-4}$.
For $n=2$ or $3$ the set of known operators is simply $\hat{Z}_0 = \hat{1}$ or $\hat{Z}_1 = \hat{v}_+$ alone, and these expansions are trivial. 

Now write a new operator,
\begin{equation} \label{eq:alphaansatz}
    \hat{Z}_n = (\hat{v}_+)^n + \sum_{k=1}^{\floor{n/2}}\alpha_{n,n-2k}\hat{Z}_{n-2k}  = (\hat{v}_+)^n + \sum_{k=1}^{\floor{n/2}}\beta_{n,n-2k} (\hat{v}_+)^{n-2k} 
\end{equation}
where $\alpha_{n,n-2k}$ are $\floor{n/2}$ undetermined constants, one for each $k$, and $\beta_{n,n-2k}$ can be written in terms of these $\alpha$s as in \cref{eq:Zeq}. 
This new operator evolves under $\mathbb{T}$ as
\begin{align*}
    \mathbb{T}[\hat{Z}_n] &= \mathbb{T}[(\hat{v}_+)^n] + \sum_{k=1}^{\floor{n/2}}\alpha_{n,n-2k}\mathbb{T}[\hat{Z}_{n-2k}] \\
    & = (1-p)\mathbb{S}_{\kappa}[(\hat{v}_+)^n] + p\,\sum_{k=0}^{\floor{n/2}}\mathcal{A}_{n,n-2k}\mathbb{C}_{\gamma}\left[:(\hat{v}_+)^{n-2k}:\right] + \sum_{k=1}^{\floor{n/2}}\alpha_{n,n-2k}\lambda_{n-2k}(p)\hat{Z}_{n-2k} \\
    & = (1-p)e^{n\kappa}(\hat{v}_+)^n + p\,\underbrace{\sum_{k=0}^{\floor{n/2}}\mathcal{A}_{n,n-2k}e^{-(n-2k)\gamma}:(\hat{v}_+)^{n-2k}:}_{\text{separate out } k=0 \text{ term, noting } \mathcal{A}_{n,n}=1} + \sum_{k=1}^{\floor{n/2}}\alpha_{n,n-2k}\lambda_{n-2k}(p)\hat{Z}_{n-2k} \\
    & = (1-p) e^{n\kappa} (\hat{v}_+)^n + p\,e^{-n\gamma} \underbrace{:(\hat{v}_+)^n:}_{\text{use \cref{eq:id1}}} +\, p\sum_{k=1}^{\floor{n/2}}\mathcal{A}_{n,n-2k} e^{-(n-2k)\gamma}:(\hat{v}_+)^{n-2k}: + \sum_{k=1}^{\floor{n/2}}\alpha_{n,n-2k}\lambda_{n-2k}(p)\hat{Z}_{n-2k} \\
    & = \underbrace{\left[(1-p) e^{n\kappa} + p\,e^{-n\gamma}\right]}_{\equiv \lambda_n(p)}(\hat{v}_+)^n + p\sum_{k=1}^{\floor{n/2}}\mathcal{A}_{n,n-2k}\left[e^{-(n-2k)\gamma}-e^{-n\gamma}\right] :(\hat{v}_+)^{n-2k}: + \sum_{k=1}^{\floor{n/2}}\alpha_{n,n-2k}\lambda_{n-2k}(p)\hat{Z}_{n-2k} \\
    & = \lambda_n(p)\hat{Z}_n + \sum_{k=1}^{\floor{n/2}}\left\{p\,\mathcal{A}_{n,n-2k} \left[e^{-(n-2k)\gamma}-e^{-n\gamma}\right] :(\hat{v}_+)^{n-2k}: + \alpha_{n,n-2k}\left[\lambda_{n-2k}(p)-\lambda_n(p)\right]\hat{Z}_{n-2k}\right\}\\
\end{align*}
We now use \cref{eq:id2} and \cref{eq:Zeq} to express all operators in the expression in brackets as powers of $\hat{v}_+$, and re-indexing and reordering the sums gives
\begin{align}
    \mathbb{T}[\hat{Z}_n] &= \lambda_n(p)\hat{Z}_n + \sum_{l=1}^{\floor{n/2}} \sum_{k=1}^{l} \Bigg\{ p\,\left[e^{-(n-2k)\gamma}-e^{-n\gamma}\right]\,\mathcal{A}_{n,n-2k}(\mathcal{A}^{-1})_{n-2k,n-2l} \nonumber \\
    & \hspace{6cm} + \left[\lambda_{n-2k}(p)-\lambda_n(p)\right]\,\alpha_{n,n-2k}\beta_{n-2k,n-2l} \Bigg\} (\hat{v}_+)^{n-2l} \\
    &\equiv \lambda_n(p)\hat{Z}_n + \sum_{k=1}^{\floor{n/2}} \mathcal{C}_{n,n-2k}\,(\hat{v}_+)^{n-2k}.
\end{align}
Demanding that all of the $\floor{n/2}$ coefficients $\mathcal{C}_{n,n-2k}$ vanish enforces particular values for the unknowns $\alpha_{n,n-2k}$.
This demand can always be satisfied because $\mathcal{C}$ can be expressed as a triangular matrix multiplying a vector of powers of $\hat{v}_+$.
For $k=1$, $\mathcal{C}_{n,n-2}=0$ gives an equation that contains only a single unknown, $\alpha_{n,n-2}$, and we find
\begin{equation}
    \alpha_{n,n-2} = \frac{p\,\left[e^{-(n-2)\gamma}-e^{-n\gamma}\right]\,\mathcal{A}_{n,n-2}(\mathcal{A}^{-1})_{n-2,n-2}}{\left[\lambda_{n-2}(p)-\lambda_n(p)\right]\,\beta_{n-2,n-2}},
\end{equation}
entirely in terms of known quantities. 
For $k=2$, $\mathcal{C}_{n,n-4} = 0$ gives an equation that depends on $\alpha_{n,n-4}$ and the now determined $\alpha_{n,n-2}$, allowing for the former to be obtained explicitly.
The pattern continues, allowing us to solve for all $\alpha_{n,n-2k}$ in sequence. 

In the end we are left with
\begin{equation}
    \mathbb{T}[\hat{Z}_n] = \lambda_n(p) \hat{Z}_n
\end{equation}
so this posited operator is a new eigenvector of $\mathbb{T}$ with eigenvalue $\lambda_n(p) = (1-p)e^{n\kappa}+p\,e^{-n\gamma}$.
We can repeat this process \textit{ad infinitum} to construct an infinite tower of eigenvectors and eigenvalues for all integers $n$.
This also confirms the consistency of the assumed form of the operators $\hat{Z}_k$ in \cref{eq:Zeq}. 
We can also rewrite \cref{eq:alphaansatz} as
\begin{equation} \label{eq:alphaZ}
    (\hat{v}_+)^n = \hat{Z}_n - \sum_{k=1}^{\floor{n/2}}\alpha_{n,n-2k}\hat{Z}_{n-2k},
\end{equation}
giving a way to express any power of $\hat{v}_+$ in terms of eigenvectors $\{\hat{Z}_k\}$.

\subsection{Evolution and control of order-$n$ operators}
Though the set $\{\hat{Z}_n\}$ does not comprise a complete set of eigenvectors of $\mathbb{T}$, it allows us to examine the controllability of generic operators. 
With repeated application of $\mathbb{T}$, the operators $\hat{Z}_n$ acquire factors of the corresponding $\lambda_n(p)$, and will diverge whenever $\lambda_n(p)>1$, allowing us to define a critical control rate for each $n$ as
\begin{equation} \label{eq:pstarSUPP}
    \lambda_n(p^\ast_n) = 1 \quad\Rightarrow\quad p^\ast_n = \frac{e^{n\kappa}-1}{e^{n\kappa} - e^{-n\gamma}},
\end{equation}
with divergence occurring for $p<p^\ast_n$ as $t\to\infty$. 
For any $\kappa,\gamma>0$ we have $p^\ast_{n+1}>p^\ast_n$, so that operators of higher order require more frequent interventions to control. 
A generic operator $\hat{\mathcal{O}}$ containing terms up to $n$th order in canonical operators can be expanded in terms of the eigenvectors of $\mathbb{T}$.
This expansion will generically contain $\hat{Z}_k$ terms for all $k\leq n$, and so will diverge at late times for all $p < p^\ast_n$.

\section{Classical noise in the control map}\label{sec:classicalnoise}

We can simulate similar features to the full quantum problem but with classical probability distributions using an Ornstein-Uhlenbeck process in phase space: $x$ evolves as
\begin{equation}
    \frac{dx}{dt} = -\gamma x + \sqrt{2D} \xi(t),
\end{equation}
where $\expval{\xi(t)\xi(s)}=\delta(t-s)$, and similarly for $p$. 
The Fokker-Planck equation for the distribution of values of $x$ (or $p$) is well known
\begin{equation}
    \frac{\partial P}{\partial t} = \gamma \frac{\partial}{\partial x}(x P) + D \frac{\partial^2 P}{\partial x^2},
\end{equation}
so the distribution evolves with the kernel
\begin{equation}
    K(x,t;x',t') = \sqrt{\frac{\gamma}{2\pi D(1- e^{-2\gamma(t-t')})}} \exp\left[ -\frac{\gamma}{2D} \frac{(x-x' e^{-\gamma(t-t')})^2}{1-e^{-2\gamma(t-t')}}\right].
\end{equation}

Taking an initial Gaussian distribution, $P(x,0) = \frac{1}{\sqrt{2\pi \sigma}} e^{-\frac{(x - x_0)^2}{2\sigma}}$, convolution with the evolution kernel yields another Gaussian
\begin{equation}
    P(x,t) = \frac{1}{\sqrt{2\pi(\sigma e^{-2\gamma t}+\frac{D}{\gamma}(1-e^{-2\gamma t}))}} \exp\left[ -\frac1{2} \frac{(x-x_0 e^{-\gamma t})^2}{\sigma e^{-2\gamma t}+\frac{D}{\gamma}(1-e^{-2\gamma t}))} \right]
\end{equation}
Therefore we can track how control changes Gaussian probability distributions
\begin{equation}
\begin{aligned}
    x & \mapsto e^{-\gamma t}x \\
    \sigma & \mapsto e^{-2\gamma t} \sigma + \tfrac{D}{\gamma}(1 - e^{-2\gamma t})
\end{aligned}
\end{equation}
This will be true for both $x$ and $p$ (or any other basis).
We also have a chaotic map which behaves the same as indicated in the main text.
This is the same as what we have for Gaussian bosonic states.

Interleaving chaos and control stochastically, we can build up a distribution of $v_\pm$ and $\sigma_\pm$. 
The controlled ``distribution'' then takes the form
\begin{equation}
    P_0(x,p) = \frac1{2\pi D/\gamma} \exp\left[-\frac{\gamma}{2D}(x^2 + p^2)\right].
\end{equation}
This form can now be integrated against another Gaussian
\begin{equation}
    P(v_+,v_-) = \frac{1}{2\pi\sqrt{\sigma_+\sigma_-}} e^{-\frac{v_+^2}{2\sigma_+} - \frac{v_-^2}{2\sigma_-}},
\end{equation}
to give
\begin{equation}
    \frac{4\pi D}{\gamma} \int dv_- dv_+ P(v_+, v_-) P_0(v_+, v_-) = \frac{2D/\gamma}{\sqrt{(D/\gamma+ \sigma_+)(D/\gamma+\sigma_-)}}.
\end{equation}
The last step is to note that we average over trajectories so there is a \emph{probability distribution} of $\sigma_{\pm}$ which we call $\mathbb{Q}(\sigma_+, \sigma_-)$. 
For this, we can compute the true probability distribution
\begin{equation}
    P_{\mathrm{full}}(x_+,x_-) = \int d\sigma_+ d\sigma_- \mathbb{Q}(\sigma_+,\sigma_-) P(x_+,x_-),
\end{equation}
and we therefore have the final expression to find the overlap with the noisy controlled distribution
\begin{equation}
    \frac{4\pi D}{\gamma} \int dx_- dx_+ P_{\mathrm{full}}(x_+, x_-) P_0(x_+, x_-) = \int d\sigma_+ d\sigma_- \frac{(2D/\gamma) \; \mathbb{Q}(\sigma_+, \sigma_-)}{\sqrt{(D/\gamma+ \sigma_+)(D/\gamma+\sigma_-)}},
\end{equation}
If we have quantum-limited noise if $D/\gamma = \frac\hbar2$.
This shows that the quantum case is related to a noisy classical problem with a particular form of noise (but \emph{only} with the control map).

\bibliography{references}